\documentclass{article}
\usepackage[left=1in, right=1in, top=1in, bottom=1in]{geometry}
\usepackage{enumitem}
\usepackage{authblk}
\usepackage{graphicx}	
\usepackage{amsmath, amsthm, amssymb}
\usepackage{xspace}
\usepackage{comment}
\usepackage{hyperref}
\usepackage{xcolor}
\usepackage{caption, subcaption}

\theoremstyle{plain}

\theoremstyle{definition}

\title{Experience Report: Standards-Based Grading at Scale in Algorithms\footnote{We thank our course staff for their hard work and support, which was instrumental in implementing SBG. We thank S. Krinsky for personal communication sharing her course structure and advice during our initial planning. We thank V. Dunn for compiling detailed grade data dating back several years. JAG thanks C. Heckmann, B. Hayes, D. Larremore, B. Waggoner, and A. Clauset for useful feedback on part of a draft. 
ML wishes to thank B.H. Payne for useful feedback on part of a draft, A. Book and A. Edwards for helpful discussions regarding the changes made in Summer '21, as well as A. Book for helpful feedback on the development of a course text \cite{LevetAlgorithmsNotes}. We thank the anonymous referees for their helpful feedback. JAG was partially supported during this work by NSF CAREER award CCF-2047756.} }

\author[1]{Lijun Chen}
\author[1,2]{Joshua A. Grochow
}
\author[1,3]{Ryan Layer}
\author[1]{Michael Levet}

\affil[1]{Department of Computer Science, University of Colorado Boulder}
\affil[2]{Department of Mathematics, University of Colorado Boulder}
\affil[3]{BioFrontiers Institute, University of Colorado Boulder}

\begin{document}
\maketitle

\begin{abstract}
We report our experiences implementing standards-based grading at scale in an Algorithms course, which serves as the terminal required CS Theory course in our department's undergraduate curriculum. The course had 200-400 students, taught by two instructors, eight graduate teaching assistants, and supported by two additional graders and several undergraduate course assistants. We highlight the role of standards-based grading in supporting our students during the COVID-19 pandemic. We conclude by detailing the successes and adjustments we would make to the course structure.
\end{abstract}

\thispagestyle{empty}

\newpage

\setcounter{page}{1}

\section{Introduction}
\label{sec:introduction}
Our department's undergraduate Algorithms course serves to meet the in-major Theory requirement. Such courses are standard in undergraduate computer science programs, though they vary substantially in content ranging from implementation-heavy courses in advanced data structures to a more theoretical and proofs-based introduction to algorithm analysis. Our course emphasizes the theoretical content, while incorporating occasional small programming assignments. 

This course presents significant and unusual pedagogical challenges, in that it aims to teach theoretical content (the algorithm design techniques and key examples), rigorous problem solving skills (applying and generalizing the techniques), and mathematical communication (i.e., mathematical proof and formalisms). The initial enrollment is also quite large, ranging from 300-400 students each semester, which limits the amount of individualized attention and detailed feedback that we can provide. The course prerequisites include some Discrete Math course, which can be fulfilled using comparable courses in the Computer Science, Math, Applied Math, or Electrical Engineering departments. The overwhelming majority of our students fulfill this prerequisite with the Computer Science department's Discrete Math course. Despite this prerequisite, Algorithms is often the first time students are expected to parse or formulate mathematical proof beyond simple numerical induction (e.g., Fibonacci, sum of the first $n$ integers). The culmination of these pedagogical challenges results in a rather steep learning curve for students. 

We contrast these challenges with Introduction to Proofs courses in Math departments, which focus on mathematical communication and adopt simpler content (e.g., logic, sets, functions, relations) as the vehicle upon which to practice formulating mathematical proofs. We also contrast our course with upper-division courses in the Math department like Abstract Algebra and Real Analysis, which require prerequisites emphasizing both content and proof-writing skills, and so focus primarily on content. In-major Math courses also tend to be much smaller than Algorithms, usually with fewer than 30 students.

Previous iterations of this course utilized more traditional points-based grading schemes, which failed to effectively communicate to students their strengths and specific areas for improvement. In addition, points-based grading factored each attempt on a given skill into student grades, which effectively punished students for erroneous initial attempts, even if the student went on to demonstrate proficiency later in the course. In our experiences, the drawbacks of points-based grading coupled with the steep learning curve created a course structure that exacerbated large-scale anxiety amongst the students. On the instructional side, implementing points-based grading was time consuming in determining how to consistently award partial credit.

In the Spring 2020 semester, we redesigned the course to incorporate \textit{standards-based grading}, which is the implementation of the principle that \textit{grades should communicate to students their current command of the content and progress with respect to well-defined standards} \cite{StangeProofs}. In addition to clearly communicating to students their command of course content, we also sought to structure the course grading scheme to reward students for their demonstrated learning prior to the end of the semester, rather than punishing students for failing to (fully) grasp a given skill earlier in the course. We refined our standards-based implementation in subsequent semesters, including Fall 2020, Summer 2021, and Fall 2021.

In this paper, we document our attempts to design an Algorithms course using standards-based grading, including successes and challenges of implementing standards-based grading at scale in a university environment. We also discuss both the challenges we faced in migrating online, as well as the flexibility that standards-based grading enabled us to provide to students. This flexibility was especially impactful during the pandemic, both because students are dealing with more physical and mental health issues, and because of the speed with which university policy changes regarding in-person vs. online vs. hybrid instruction.

\subsection{Standards-Based Grading}

Standards-based grading is a scheme in which students are provided a clearly defined list of learning objectives and multiple attempts to demonstrate proficiency. While ideally students persist until they have demonstrated proficiency for each topic, we note that this is not a requirement for standards-based grading \cite{StangeProofs}. Final course grades are based upon the number of standards for which students have demonstrated proficiency and the degree of demonstrated learning for each standard \cite{Beatty2013StandardsbasedGI, Duker2015HackingTM, StangeProofs, Elsinger2019ApplyingAS, Lewis2020GenderEO}. 

Standards based grading is a popular topic in K-12 education \cite{marzano_2011, reeves_2008, vatterott_2015}, dating back to K-12 educational reform in the 1990s \cite{marzano_2011}. It has begun to take root in the university setting, most notably in lower-division math \cite{owens_2015, StangeProofs, Elsinger2019ApplyingAS, Lewis2020GenderEO} and science \cite{Beatty2013StandardsbasedGI, Post2017StandardsBasedGI} courses. However, standards-based grading has also been adopted in upper division math courses such as Real Analysis and Abstract Algebra \cite{SalomoneOrigins, SalomoneImplementation}, as well as other disciplines such as educational leadership \cite{Townsley2019WalkingTT, Buckmiller2017QuestioningPA} and music \cite{Duker2015HackingTM}. 

We are only aware of one other instance where standards-based grading has been adopted in Algorithms \cite{ShindlerAlgorithmsMBG}. Here, the authors organized the content into five core topics: dynamic programming, greedy algorithms, divide-and-conquer, network flows, and computational complexity. Students can earn credit for these topics using periodic assessments throughout the semester, as well as the final exam. In particular, if students demonstrate understanding on a topic early the semester, then they need not attempt that topic on the final exam. The authors observed that this approach resulted in significant improvement in student performance for dynamic programming and proof-writing, which are two of the more challenging topics in the course. Additionally, over two-thirds of students increased their grade by at least 5\% by taking the final exam.

Faculty that have implemented standards-based grading report a number of key benefits, both for students and instructors. In particular, students have reported that standards-based grading helped them to focus on understanding the content rather than accruing points \cite{Buckmiller2017QuestioningPA, Elsinger2019ApplyingAS}. Math instructors have also observed reduced student stress in courses utilizing standards-based grading \cite{Elsinger2019ApplyingAS, Lewis2020GenderEO, SelbachAllen2020RaisingTB}. On the other hand, student anxiety and discomfort may be initially exacerbated by standards-based grading due to familiarity with more traditional points-based grading systems \cite{Post2017StandardsBasedGI}. Helping students understand the reassessment process is key to quell this anxiety and increase student buy-in \cite{Elsinger2019ApplyingAS}.

\section{Course Content}

We begin by detailing the course content, including details about how the material was presented and the order in which the topics were covered. In subsequent iterations of the course (Summer 2021 onwards), we adjusted both the manner and order in which we covered certain topics in order to help students gain traction, provide better feedback, and reduce the grading workload. 

The key topics were as follows (for a current treatment, see \cite{LevetAlgorithmsNotes}).
\begin{itemize}
\item \textbf{Proof by Induction \& Loop Invariant Proofs}.
\item \textbf{Asymptotics}. Comparing functions (Calculus techniques), Unrolling and Tree Methods for finding tight asymptotic bounds on recurrences.
\item \textbf{Analyzing Code}. The goal was to write down the runtime complexity function of a given algorithm, such as iterative algorithms where the loop variables were independent or dependent. For recursive algorithms, students were asked to write down a recurrence for the runtime complexity.

\item \textbf{Path-Finding}. Breadth-first search, Depth-first search, Dijkstra's Algorithm.
\item \textbf{Spanning Trees}. Safe \& Useless Edges, Prim's Algorithm, Kruskal's Algorithm.
\item \textbf{Ford--Fulkerson}. Here we emphasized the residual graph, but eventually switched to using the flow network (see Sec.~\ref{sec:second}).
%Here, we emphasized the residual graph, which contained the same vertex set as the initial flow network. There was a directed edge $(i, j)$ if positive flow could be pushed from $i \to j$. The edge $(i, j)$ was labeled with the amount of flow that could be pushed from $i \to j$ direction. 

\item \textbf{Reductions to Max Flow.} Students were asked to construct or reason about reductions to the Max Flow problem (e.g., Bipartite Matching).

\item \textbf{Greedy Algorithm Principles}. Examples where greedy algorithms fail to yield optimal solutions (often related to the Interval Scheduling and Making Change problems), Exchange Arguments.

\item \textbf{Divide \& Conquer}. Examples where a divide \& conquer algorithm failed to return the correct result, Quicksort, Mergesort
\item \textbf{Dynamic Programming}.
\item \textbf{P vs. NP Problem}. Showing problems belong to P or NP; Structure and Consequences of the P vs. NP problem.
\end{itemize}

Prior to Summer 2021, we covered the course content in the order above, with the first half of the course notably covering proof by induction \& loop invariant proofs, and asymptotics. Despite the fact that the key skills (proof by induction, exponentials \& logarithms, calculus techniques) are prerequisite content, students report that they find this material to be quite technical. In particular, due to poor early performance on these topics, many students reported that they felt set up to fail in the course. 

Our treatment of the Ford--Fulkerson algorithm using residual graphs was a pressure point for both students and grading staff. Namely, constructing and updating the residual graph at each iteration of the Ford--Fulkerson algorithm was quite involved. As a result, students frequently made minor careless errors (e.g., omitting an edge) which had significant impact on the final answer. Additionally, grading student work (even when requiring an initial flow-augmenting path) was quite tedious, with our graders reporting spending 3-5 minutes per submission.

\section{First Attempt: Standards-Based Grading (Spring and Fall 2020)} \label{SBGAttempt1}

In the Spring and Fall 2020 semesters, we implemented standards-based grading, based largely on the logistics similar to those of Heubrach \& Krinsky \cite{Heubach2020ImplementingMG}. % We should cite personal communication with S. Krinsky here when it gets de-anonymized
We began by identifying 25 content standards, outlined as follows.
\begin{itemize} 
\item \textbf{Proof by Induction \& Loop Invariant Proofs (1 standard)}. 
\item \textbf{Asymptotics (3 standards).} Calculus Techniques, Unrolling, Tree Method. 
\item \textbf{Analyzing Code (3 standards)}. Nested independent loops, Nested dependent loops, Recursive algorithms. 

\item \textbf{Path-Finding (2 standards)}. Breadth-first search/Depth-first search, Dijkstra's Algorithm.
\item \textbf{Spanning Trees (2 standards)}. Safe \& Useless Edges, Prim's \& Kruskal's Algorithms.

\item \textbf{Network Flows (2 standard)}. Ford--Fulkerson, Reductions to Max Flow.

\item \textbf{Greedy Algorithm Principles (2 standards).} Examples where greedy algorithms fail to yield optimal solutions, Exchange Arguments.

\item \textbf{Hashing (1 standard).}
\item \textbf{Divide \& Conquer (2 standards)}. Examples where a divide \& conquer algorithm failed to return the correct result, Quicksort analysis.

\item \textbf{Dynamic Programming (5 standards).} Identifying subproblems, Writing down recurrence, Designing DP algorithms with one-dimensional lookup table, Designing DP algorithms with two-dimensional lookup tables, Order of subproblems \& backtracking.

\item \textbf{P vs. NP Problem (1 standard).} 

\item \textbf{Participation standard.} 
\end{itemize}

Each content standard had four attempts: a homework, a recitation quiz, a midterm, and the final exam. In order to earn credit for a standard, students had to demonstrate proficiency twice for that standard. Students also had three reassessment attempts via a timed quiz format, provided they first wrote corrections and reflections for their previous attempts (or a correct attempt and explained the material, if the student did not attempt the given standard). Effectively, students had to demonstrate proficiency at least once on a timed assessment. A number of the standards focused on mechanical problems, such as solving a recurrence using the unrolling technique or applying Dijkstra's algorithm to a given graph. Other standards emphasized tasks that required students to generalize concepts or apply techniques in new ways, such as applying a max-flow algorithm to solve new problems, and grappling with the consequences of NP-completeness. 

We also included a Participation standard, which was designed to incentivize students to regularly engage with the content rather than relying solely on the midterms and final exam. Our initial intent was to use the Participation standard to incentivize students to regularly attend weekly recitations and work through online multiple choice mini-quizzes, which were different than the recitation quizzes. When we migrated the course to an online format in light of COVID-19, we significantly relaxed the threshold for the Participation standard. Essentially, the only students who did \emph{not} earn credit for Participation are those who didn't engage with the course at all, even before the shift to remote instruction.

Final grades correlated to the number of standards for which students demonstrated proficiency twice, with each standard receiving equal weight. Students did not receive partial credit for a standard if they demonstrated proficiency only once. In order to pass the course (with a C-), students needed to demonstrate proficiency twice on 14 standards. Each additional standard they accrued corresponded to an additional one-third of a letter grade. The cutoff for an A was 21 standards.

Individual problems were graded using five categories: No Attempt, Significant Errors, Progress, Proficiency, Near Perfect. We colloquially referred to these numerically as 0--4, but  repeatedly emphasized to students their categorical meaning. For brevity we refer to them here numerically. 
%Here, we used a 3 to indicate that a student had mastered the concept, but that their solution had room for improvement. A grade of 4 corresponded to a near-perfect solution. 
Students who earned below a 3 on a problem did not receive any credit on a problem. Scores below a 3 were used to provide feedback to students as to their progress on the problem, in addition to written feedback. % Here, a 0 corresponded to no attempt, a 1 indicated that a student made an attempt, and a 2 indicated that the student made reasonable progress but did not demonstrate mastery. Effectively, the only threshold that impacted student grades was whether they received at least a 3. In particular, 
Students did not receive any additional credit for earning a 4 rather than a 3. Rather, we designed the 3 vs. 4 distinction to recognize exceptional student solutions, as well as to help alleviate student anxiety by signaling a difference between proficiency and near-perfection.

Quizzes and exams each only had one problem per standard. So a student's proficiency score on that question factored directly into their overall course grade. An individual homework assignment, however, often included multiple questions covering a given standard. These scores were then aggregated to determine the overall proficiency score. Students who received at least a 3 on each question for that standard received a 3 or 4 for their overall proficiency score, while students who received below a 3 on every question for that standard received at most a 2 for their overall proficiency score. Similarly, students who received a grade of 0 or 1 on any question for a given standard received at most a 2 for their overall proficiency score. In borderline cases, where students had a mix of 2's, 3's and 4's, we examined their work more closely. Students who demonstrated proficiency on the more challenging questions and had a preponderance of 3's and 4's received an overall proficiency score of at least 3. On the other hand, students who struggled on the more challenging questions or earned primarily scores of 2 received at most a 2 for their overall proficiency score. %In exceptionally borderline cases, we looked more closely at their individual solutions to ascertain mastery. 

\textbf{What Worked Well.} Much of what worked well in Spring and Fall 2020 continued to work well in subsequent iterations of the course. We defer this discussion to Section \ref{Discussion}.

\textbf{Pandemic Teaching.} The transition to online teaching in the middle of the Spring 2020 semester, associated with the COVID-19 pandemic, necessitated structural changes in the course to accommodate the online format, as well as to provide flexibility and support to students. Prior to the pandemic, students had weekly homework, a 25-minute closed-book quiz in recitation each week, and a weekly online mini-quiz that contributed to the Participation standard. Homework was to be submitted directly to Canvas, and so little changed in that regard. Each weekly recitation quiz covered anywhere between one to three standards, with one question per standard. 

After moving online, each weekly recitation quiz was broken up so that each question was its own timed Canvas assignment, which students could take at any time over a four day period. This allowed students more flexibility to complete the quizzes at their convenience. We calibrated the timing to allow 15 minutes to take the quiz, scaled for students with extra time accommodations, and an extra 15 minutes for students to upload their work to Canvas. In practice, students could use some of the time allocated for uploading work, in order to continue answering the question. Students reported that the online quiz format decreased the pressures associated with a 25-minute in-person quiz. Starting in Fall 2020, students were given 45 minutes per quiz, with an intended 30 minutes to take the quiz and 15 minutes to upload their work to Canvas. We continued this practice even in Fall 2021, when the class had both in-person and online sections, because of the ease of logistics.

Students were permitted to use their notes and book, but collaborating with other students and using online tutoring services such as Chegg were prohibited. We also continued requiring students to show their work and document their reasoning. Requiring work and timing the quizzes mitigated the usefulness of posting to message boards or tutoring services, and also allowed us to better ascertain student understanding.

\textbf{Participation Standard.} After moving online in Spring 2020, the criterion for earning the Participation standard were opaque. This inspired significant student anxiety and required a substantial amount of time on the parts of the instructional staff to manage. During the Fall 2020 semester, the criterion for the Participation standard centered around recitation attendance. As recitations were remote, it incentivized students to log in for attendance credit without engaging in group work activities. Eventually, because of the ongoing pandemic, we removed Participation standards and adjusted the grade scale accordingly.

%Moving forward, we would recommend calibrating the Participation standard solely to recitation or removing it altogether. In addition to attending the duration of recitation, we would also implement a short exit ticket for students to reflect upon the given lesson. Students would earn credit towards the Participation standard for both reasonably thoughtful responses and attendance. The Participation standard would then be awarded to students who earned credit for a certain number of recitations.

%We advocate against using attendance as the only criterion for awarding the Participation standard, as it incentivizes students to attend without being engaged. This is a particular issue over Zoom, when it is easier for students to  The exit ticket requirement has an additional advantage, in that it provides a means of formative assessment. This is particularly valuable to help the course instructors identify student misconceptions. As the instructors take a minimal role in grading, it can be difficult to identify student misconceptions and common mistakes. The responses from the exit tickets would allow us to modify lesson plans and design homework problems to address such misconceptions. \\

%In large classes, it may be worthwhile to revisit the number of times students are required to demonstrate mastery to earn credit for a given standard. We required students to demonstrate mastery twice, to help ensure that they retained the concept and make it harder to cheat. Requiring only one attempt at mastery would significantly reduce the workload for the instructional staff. 

\textbf{Student Buy-In.} A key challenge we faced was in getting student buy-in to the grading scheme. Algorithms is a course in which many students struggle to gain traction. This was exacerbated by the fact that the course has traditionally been front-loaded with respect to the material students find most challenging (i.e., proof by induction, asymptotics). Students also expressed that the Calculus standard covered too much material. Combined with the technical nature of this material, students found attempting this standard to be  particularly frustrating. We note that regardless of whether the course employed traditional points-based grading or standards-based grading, student performance early in the semester reflected these difficulties. 

With standards-based grading, our goal was to incentivize students to work out the problems completely and correctly. While minor mistakes would make the difference between a 3 and a 4 (both of which counted for full credit), more serious mistakes could result in grades of 1 or 2 for individual problems. While the thresholds for earning a 1 or 2 were quite different, neither score counted towards proficiency. For this reason, students perceived that there was no partial credit under this grading scheme. This observation, coupled with early poor performance, led students to quickly believe they were doomed to failure. 

For problem sets, there were often multiple problems for a given standard. Precisely, if there were $d$ problems for a given standard, we defined a mapping $f : \{0, 1, 2, 3, 4\}^{d} \to \{0, 1, 2, 3, 4\}$ to compute the overall standard score on that assignment. Only the overall standard score was what impacted students' final grades. This mapping was often not a numerical formula; and in particular, rarely the average. Additionally, we did not share any information about our mapping with students, other than that we did not in general use the average to compute overall standard scores. As a result, students struggled to reconcile their scores on each individual problem with their overall standard score. This led to significant confusion and frustration for the students.

After the first midterm, when students began demonstrating proficiency more readily (even on new material), the perception began to change. By the end of the semester, most students earned an A or B in the course, and we received many \emph{strongly} positive comments about the grading scheme (including several students wishing that all their courses would adopt such a scheme).

%In order to increase student buy-in, as well as to help students gain traction and build confidence, we would incorporate more opportunities for early victories. In particular, the fourth author would recommend placing greedy algorithms before the material on asymptotics. Historically, students find this content to be more intuitive and motivated, compared to the material on asymptotics. In particular, our treatment of greedy algorithms is also fairly introductory. While students are expected to prove small lemmas using the exchange argument technique, we do not expect them to arrive at algorithm correctness proofs on problem sets or exams. Additionally, the fourth author recommends sharing the cutoffs for an overall standard score of 3 (full credit) in order to reduce student confusion and frustration.

\section{Second Attempt: Standards-Based Grading (Summer \& Fall 2021)} \label{sec:second}

During Summer and Fall 2021, JAG and ML restructured the course to increase student buy-in. The two key changes involved reordering the content and restructuring the standards. The course structure was otherwise identical as in Spring and Fall 2020.

We begin by detailing changes we made to the standards.

\begin{itemize} 
\item \textbf{Proof by Induction}. We removed Loop Invariant Proofs from the course. We note that students already struggle with Proof by Induction, and it is a more transferable skill to other courses. Additionally, students found Loop Invariant Proofs to be unmotivated. 

\item \textbf{Asymptotics.} We broke the Calculus standard into two separate standards: Calculus I Techniques (L'Hopital's Rule), and Calculus II Techniques (Ratio \& Root Tests). As a result, our extensive problem set on this could be naturally divided into two smaller sections. This better conveyed to students their understanding of the respective techniques, while also reducing frustration at lack of partial credit.

\item \textbf{Spanning Trees}. We broke Prim's and Kruskal's algorithms into two standards, to build in early victories for students. 

\item \textbf{Network Flows}. Due to the technical nature of residual graphs, we moved to working with the flow networks. We added a standard on Network Flow Terminology to ensure that students were comfortable reading a flow network and finding flow-augmenting paths. We also kept our Ford--Fulkerson standard, asking students to execute the algorithm on the flow network rather than using the residual graph construction. As a result, grading time for each of these standards was reported to be 20 seconds per problem, and over 90\% of the class demonstrated proficiency twice for both these standards by the end of both the Summer and Fall 2021 offerings.
	
\item \textbf{Dynamic Programming.} We broke Dynamic Programming up into four key standards: Identifying subproblems, Writing down recurrence, Using the recurrence to fill in a lookup table, and Designing DP algorithms. The key techniques we stressed involved identifying the recursive structure (formalized with writing down a recurrence), and then using that recursive structure to construct and fill in an appropriate lookup table. 

The final standard here, Designing DP algorithms, was only assessed twice via untimed problem sets. Students only had to demonstrate proficiency once in order to obtain credit for this standard. Because this standard was really a synthesis of several other standards, we felt that giving students time to think here was better for their learning, as opposed to testing this standard on a timed assessment. 

\item \textbf{P vs. NP Problem.} We broke the P vs. NP standard up into four distinct standards: Formulating decision problems, Showing a problem belongs to P, Showing a problem belongs to NP, and Structure \& Consequences of P vs. NP. This allowed us to better communicate key skills to students, as well as to use the Structure \& Consequences of P vs. NP standard as an opportunity to challenge our students. Sample problems under the Structure \& Consequences standard include asking students to recognize when someone is reducing in the wrong direction (a common mistake in NP-completeness proofs), the fact that the RSA cryptosystem can be broken if P = NP, and asking students to fill in the details of a proof showing that $\text{NP} \subseteq \text{PSPACE}$.\footnote{Complete writeups of these problems are available by request.} 
\end{itemize}

As a result, there were 30 content standards (plus the Engagement standard in Summer 2021). Each core topic also had a mix of mechanical standards (e.g., work through an algorithm, come up with a counter-example), as well as more involved synthesis and proofs-based standards. There were 23 mechanical standards, as well as the Engagement standard (more below). The cutoff for a C- was 23 standards in Summer 2021, and 20 standards in Fall 2021. In order to earn a B- or higher, it was necessary for students to demonstrate proficiency twice for a subset of the standards emphasizing synthesis or proofs-based problem solving.

\textbf{Reordering Content.} Aside from revising the standards, the key curricular change effectively involved moving greedy algorithms to the front of the course. In particular, we covered Proof by Induction, Pathfinding, Spanning Trees, and Network Flows (including Reductions to Max Flow) for the first midterm. Unlike in previous semesters, we did not receive complaints (let alone, organized complaints) from students as to the difficulty of the course. During the Summer 2021 session, several students who were repeating the course praised the change in the content ordering for helping the class to gain traction. At least one student reported to ML that the previous course ordering felt like ``having a brick thrown in one's face." 
	
The grades after the first midterm reflected the positive changes. In Summer 2021, the first quartile cutoff indicated that students had demonstrated proficiency twice for 6 standards and \textit{at least} once for 10 standards. In Fall 2021, the first quartile cutoff indicated students had demonstrated proficiency twice for 6 standards and \textit{at least} once for 9 standards. We contrast this with the Fall 2020 offering, where the first quartile cutoff students had demonstrated proficiency twice for 2 standards and \textit{at least} once for 7 standards.

One concern in reordering the content is that it would make place the majority of the more difficult content (asymptotics, dynamic programming, and P vs. NP) in the second half of the course, possibly setting students up to fail. Despite these changes, we actually saw an increase in the number of students that passed the course with a grade of at least C- (we only count students who made it to the end of the course). We had pass rates of 173/222 ($77.93\%$) in Fall 2020, 51/64 ($79.68$\%) in Summer 2021, and 199/253 ($78.65\%$) in Fall 2021. We also note that in Fall 2020 and Fall 2021, we began the semester with approximately 270 students. So not only did our pass rate increase from Fall 2020 to Fall 2021, but the number of students who dropped also decreased.

\textbf{Participation/Engagement Standard.} For Summer 2021, the Participation standard was renamed to be the Engagement standard. In order to earn credit for the Engagement standard, students had to complete a syllabus quiz and engage in at least 5/7 recitations. Each recitation was held in a remote synchronous format, with the TA facilitating active learning and group work. At the end of recitation, students submitted responses to a virtual exit ticket explaining in 1-2 sentences a concept they learned, as well as asking a question or identifying a confusing concept about the material from that day. Students who attended recitation and submitted a reasonably thoughtful response received credit for that week. Effectively, a thoughtful response required students to reflect, explain, or ask about a topic that was not definitional. For instance, \textit{I learned about bipartite graphs} is not a thoughtful response, as a non-engaged student could simply look at the worksheet for that week. Instead, they would need to explain something they learned about bipartite graphs; for instance, \textit{I learned that bipartite graphs have no cycles of odd length. Therefore all trees are bipartite.} 

Students who attended recitation were generally engaged, though there were still some folks who did not discuss with their groups or ask questions. For students who did not wish to discuss in recitation, the exit ticket served as a means to assess their engagement. With rare exceptions, student responses on the exit tickets were quite thoughtful. The TAs also observed that students generally learned quite a bit in recitation, even if they struggled. There were 46/64 students who earned credit for the Engagement standard; of these, 41 (89.1\%) students earned a passing grade. We contrast this with the pass rate of 51/64 (79.68\%) students. While our goal was not to assess the benefits of active learning, it is worth noting that the gains amongst these students appear consistent with existing evidence on the effectiveness of active learning in increasing student learning and performance \cite{Theobald6476, Freeman8410, McConnellActiveLearning, YoderHochevar}.  

Due to the logistics of having in-person and remote recitations in Fall 2021, we did not include a Participation or Engagement standard.

\section{Discussion} \label{Discussion}

\noindent \textbf{What Worked Well.} Our standards-based grading scheme had several key advantages. When grading, the course staff were able to focus on providing feedback rather than determining appropriate quantities of partial credit. This allowed us to grade more quickly. Additionally, in instances where multiple graders were assigned to the same problem, we observed increased consistency in grading compared to previous semesters, though grading consistency remains an issue at this scale.

Standards-based grading also enabled us to clearly communicate to students their understanding of the concepts and techniques. This reduced student anxiety around grades, in that students could readily identify specific standards on which to focus, based on their desired end of semester grades. Furthermore, standards-based grading allowed students to focus on growth without being hampered down by low grades resulting from earlier assignments that fell short of expectations. In addition, students were less likely to dispute their grades; and those that did dispute their grades were less apt to negotiate for points on individual problems. Conversations in office hours centered around the course content rather than individual student standing. % they had mastered, as well as those where they needed additional work. 

%Furthermore, standards-based grading provided a great deal of flexibility for how students engaged in the course, including which standards students sought to master and the timeframe in which students could master these standards. In particular, using 

\textbf{Online Quizzes.} At the onset of the COVID-19 pandemic, we migrated our quizzes to Canvas. We kept this format in subsequent semesters, in part due to the class being held in either a remote or hybrid format. The online format of the quizzes affords a number of advantages that make it worth permanently retaining. First, the instructional staff are responsible for providing accommodations to students with 1.5x testing time. For in-person quizzes, we must schedule a time to proctor these students, which is a logistical difficulty. Students who miss class also often want to make up the in-person quizzes, which creates an additional logistical burden. Using Canvas allowed us to easily grant individual students extra testing time or extended deadlines. With initial enrollments of approximately 300-400 students, this was a significant convenience.

\textbf{Reassessment Tokens.} The reassessment mechanism greatly contributed to student learning and reducing student anxiety. Student exposition on the first round of corrections \& reflections often demonstrated a significantly stronger understanding of the content than the original problem set or quiz submissions. For students whose first correction \& reflection submissions needed additional work, the additional discussion with the instructional staff served to largely clarify remaining misconceptions. The revised correction \& reflection submissions reflected this. The overwhelming majority of students demonstrated proficiency on the reassessment quizzes. In cases where students had extenuating circumstances, we granted free additional retake tokens. This allowed students to focus on their circumstances in the short-term and attempt the material at a later point, when they were able to better devote the time.

Developing the reassessment quizzes- particularly for our more involved standards like Exchange Arguments and Reductions to Max Flow- and processing student corrections \& reflections resulted in a significant undertaking. We received close to 380 submissions during the semester, with roughly 225 of those in the last 24 hours. We allocated two TAs to process these, so that students could take their reassessment quizzes and have grades back within 24 hours of the final opening. As most students wait until the last 24 hours to submit their corrections \& reflections, we recommend opening it roughly six weeks before the end of the semester and closing it at least 1.5 weeks before the final. This minimizes the need to create multiple reassessment quizzes for a given standard, which is particularly helpful for the more involved standards.

\textbf{Participation/Engagement Standard.} The Participation standard incentivized students to attend recitations, though without necessarily engaging with any active learning activities. The exit ticket requirement introduced in Summer 2021 successfully incentivized engagement, which correllated with a higher pass rate amongst the students that regularly attended recitation. While tracking exit tickets was feasible in a smaller summer class with remote synchronous recitations, it would not have been in a larger Fall or Spring offering. For these reasons, we have discontinued the Participation/Engagement standard. 

\textbf{Instructional Workload.} The two key challenges centered around developing course infrastructure (problem sets, quizzes, solutions), as well as the additional grading load incurred by the weekly quizzes. This was particularly problematic in Spring 2020, when we developed the course from scratch. In subsequent semesters, we reused existing infrastructure, though a significant amount was still developed from scratch in Fall 2020 and Summer 2021. We recommend developing the problem set and quizzes for a given standard at once.

Our TAs also spent almost all of their allotted hours on grading, which left little time for developing course infrastructure and other logistical tasks. The situation greatly improved once we hired graders, though the hiring process took several weeks. The grading situation improved in subsequent semesters, as we requested and received two additional graders.

\section{Student Performance} \label{Conclusion}
Our goal in this section is to analyze student enrollment and performance trends, comparing the Fall 2019-2021 semesters.  In order to contextualize this information, we begin by discussing the structure of our undergraduate degree programs and its relation to both the Algorithms course and the student population.

There are two undergraduate CS degrees: one by admissions in the College of Engineering and Applied Sciences (BS), and one open enrollment in the College of Arts \& Sciences (BA). The BS degree was the University of Colorado Boulder's first CS degree. The BA degree launched in 2013, with the intent to provide flexibility for students to obtain relevant CS skills to pair with their related discipline (such as for a double major). As such, Algorithms witnessed its first BA student enrollments during the Fall 2016 semester, with 59/146 students. 

While core classes such as Algorithms and Programming Languages have always been required for the BS degre, they were not initially required for the BA degree. Functionally, the department observed that students were using the BA degree either as a software engineering degree or to avoid these more challenging core classes. For this reason, the department changed the requirements for the BA degree to include both the Algorithms and Programming Languages courses. These changes became effective in the Fall 2019 semester.

We also note that the class population for Algorithms is significantly different between Spring and Fall semesters. In particular, Algorithms is scripted to be taken in the Spring semester of the third year. Thus, Spring semester tends to have a stronger student population. In the Fall semester, the population is more bimodal, with a significant number of students who are either repeating the course or who are in their second year (and thus, taking Algorithms ahead of schedule). For this reason, we report student performance separated out by Fall/Spring. Additionally, we note that our results can most closely be compared to the Fall 2019 semester. 

We have yet to establish a solid baseline to measure the success of our standards-based grading implementation for Spring iterations of the course. In Spring 2020, most of the standards were largely mechanical in nature. Additionally, with the onset of the COVID-19 pandemic, we reduced the demand of our weekly problem sets out of necessity for both the students and instructional staff. As a result, we expect that student performance in Spring 2020 to be higher than under our updated curriculum from Summer 2021 and Fall 2021. During the Spring 2021 semester, none of the authors of this paper were involved with the Algorithms course. While we are aware that the instructors for Spring 2021 used traditional points-based grading, we do not have sufficient insight into the course structure to compare it to our standards-based grading implementation or re-ordered curriculum. 

\begin{figure}[!htbp]
\centering
\begin{subfigure}[b]{\textwidth}
\centering
\includegraphics[scale=0.88]{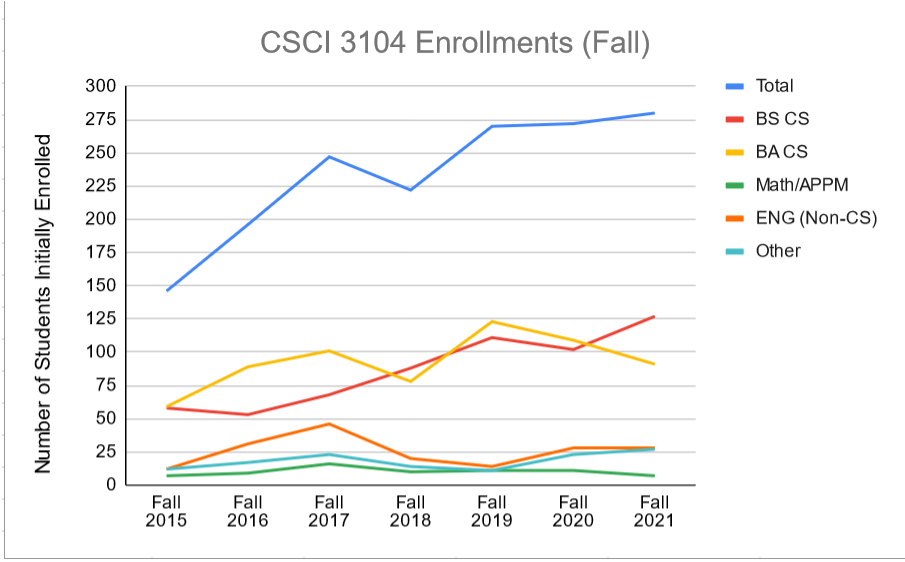}
%\caption{\label{fig:enroll_fall} Enrollments for the Fall semester in CSCI 3104 Algorithms.}
\end{subfigure}

\begin{subfigure}[b]{\textwidth}
\centering
\includegraphics[scale=0.88]{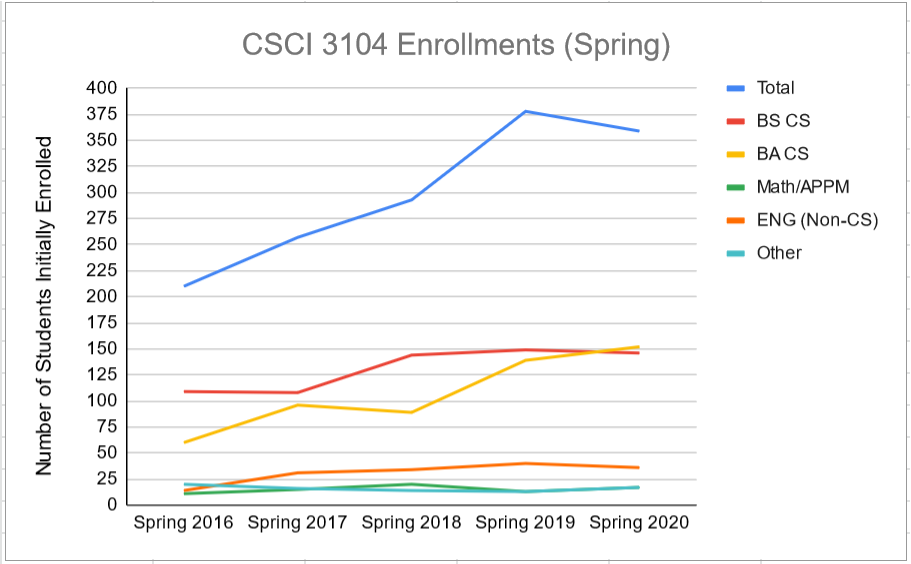}
%\caption{\label{fig:enroll_spring} Enrollments for the Spring semester in CSCI 3104 Algorithms.}
\end{subfigure}

\caption{\label{fig:enroll} Enrollments for CSCI 3104 Algorithms by major. Above: Fall semesters; below: Spring semesters.}
\end{figure}

%\begin{figure}[!htbp]
%\begin{center}
%\includegraphics[scale=0.88]{Enrollments_Spring.png}
%\caption{\label{fig:enroll_spring} Enrollments for the Spring semester in CSCI 3104 Algorithms.}
%\end{center}
%\end{figure}
We begin by detailing our enrollments, separated by the Fall and Spring, as well as by major: BS CS, BA CS, Math \& Applied Math (APPM), Non-CS Engineering, and Other; see Figure~\ref{fig:enroll}. Enrollments for both the BS CS and BA CS majors has increased steadily during both the Fall and Spring semesters, while enrollments for Math/APPM, Non-CS Engineering, and Other majors have remained steady at under 50 students. Due to the fact that Algorithms is not required for non-CS majors and the comparatively small enrollments of these students, we restrict attention to analyzing performance for the entire class, as well as the BA and BS CS populations.

We note that the demand for the CS major has exploded \cite{CRAEnrollments, BoomingEnrollments}. Our own department has witnessed this growth. Teaching resources (e.g., instructor and TA lines, funding for additional graders) have not scaled accordingly, which is a key difficulty in adequately supporting our students. For Fall iterations of Algorithms from 2016--2019, there is a clear decrease in performance for all of our students (for more detailed analysis, see Section \ref{PracticalPerformance}). Thus, we observe that student performance is inversely correlated to increased enrollments. While growing enrollments have resulted in decreased performance across all student populations in Algorithms, this change has most negatively impacted our BA CS students who (as a population) have notably lower performance metrics in comparison to students in other majors.

Our department's Discrete Math course, which is a key prerequisite for Algorithms, faces similar systemic barriers as Algorithms, in enrollment sizes and multiple learning outcomes (e.g., content, synthesis-based problem solving, and mathematical proof \& formalism). However, while Algorithms receives significant TA support, the Discrete Math course does not receive any TA lines. Given the fact that Algorithms has witnessed decreased student performance corresponding to increased enrollments, it is natural to ask whether this trend is occurring in Discrete Math. Such a trend would likely be a contributing factor to the decreased student performance in Algorithms witnessed from Fall 2015-Fall 2019. In particular, we note that the data about performance in Algorithms for the BA CS student population is consistent with departmental observations in required prerequisite CS courses. For these reasons, we conjecture that, as enrollments have grown, the lack of  TA support in the department's Discrete Math course has disproportionately left the BA CS students without sufficient scaffolding and support to thrive when they reach Algorithms. Enrollments for the department's Discrete Math course have remained steady over the last three years at approximately 300 students per semester, managed by 1-2 instructors, 2-3 graders, and additional undergraduate course assistants to hold office hours. There is also no recitation for the Discrete Math course. As a result, it would be a Herculean task to implement known effective strategies such as active learning \cite{Theobald6476, Freeman8410, McConnellActiveLearning, YoderHochevar}, frequent low-stakes assessment \cite{GuvenTesting, ROEDIGERIII20111, KarpickleBlunt, ROEDIGER201120}, and detailed feedback \cite{RazzaqFeedback, StovnerFeedback, HattieTimperley, STOVNER2022103593}. We also believe that, due to systemic barriers such as large enrollments and lack of adequate teaching resources, that it is not feasible to provide students with adequate practice \textit{and feedback} for proof-writing and higher-level synthesis-based problem solving. In our opinion, practice and feedback with such skills is key to preparing students for subsequent CS Theory courses like Algorithms.

We consider three measures of student performance, including the pass rate, retention rate, and GPA. The pass rate considers the fraction of students that passed the course (with a grade of at least C-), out of those who made it to the end of the course. The retention rate considers the fraction of students that passed the course (again, with a grade of at least C-), out of those who initially attempted the course.\footnote{Finer-grained FERPA-compliant data of student performance is available upon request.}

We first observe that from Fall 2015 through Fall 2019 (including Fall and Spring semesters), student performance across all three metrics (pass rate, retention rate, and GPA) has steadily declined. This drop in performance quite stark amongst the BA CS population, particularly in the Fall 2019 semester when CSCI 3104 Algorithms became required for the BA major. For the BA CS majors in Fall 2019 (see Table~\ref{table:perf}), the pass rate was 55.56\%, the retention rate was 44.72\%, and the GPA was 1.49. This contrasts with the course overall pass rate of 70.56\%, retention rate of 60.37\%, and GPA of 2.16. For the BS CS majors in Fall 2019, the pass rate was 83.67\%, the retention rate was 73.87\%, and the GPA was 2.74.

\begin{table}[htp!]
\begin{center}
\begin{tabular}{|r|c|c|c|}
\hline 
 & Overall & BS CS & BA CS  \\ \hline
Retention Rate & 70.56\% & 83.67\% & 44.72\% \\ \hline
Pass Rate & 60.37\% & 73.87\% & 55.56\% \\ \hline
GPA & 2.16 & 2.74 & 1.49 \\ \hline
\end{tabular}
\end{center}
\caption{ \label{table:perf} Course Performance in Algorithms for Fall 2019.}
\end{table}

\newpage
\subsection{Student Performance (Administrative Summary)}

\begin{figure}[!htbp]
\centering
\begin{subfigure}[b]{0.45\textwidth}
\includegraphics[scale=0.44]{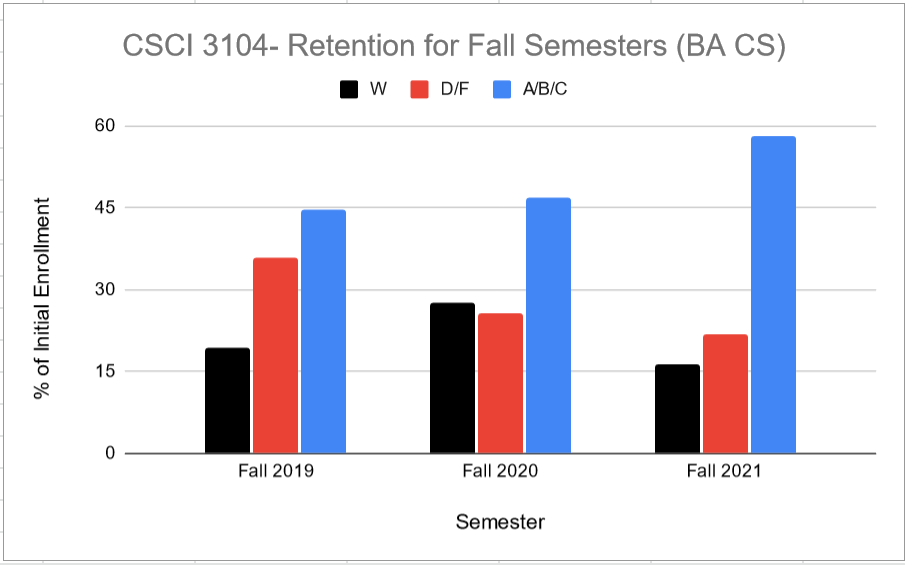} 
\caption{}
\end{subfigure}
\hfill
\begin{subfigure}[b]{0.45\textwidth}
\includegraphics[scale=0.44]{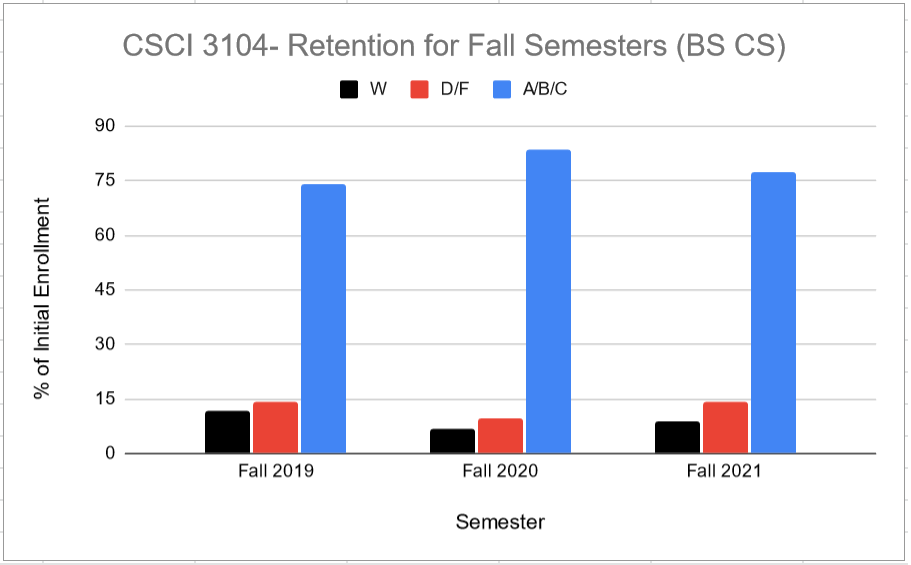} 
\caption{}
\end{subfigure}

\begin{subfigure}[b]{\textwidth}
\centering
\includegraphics[scale=0.44]{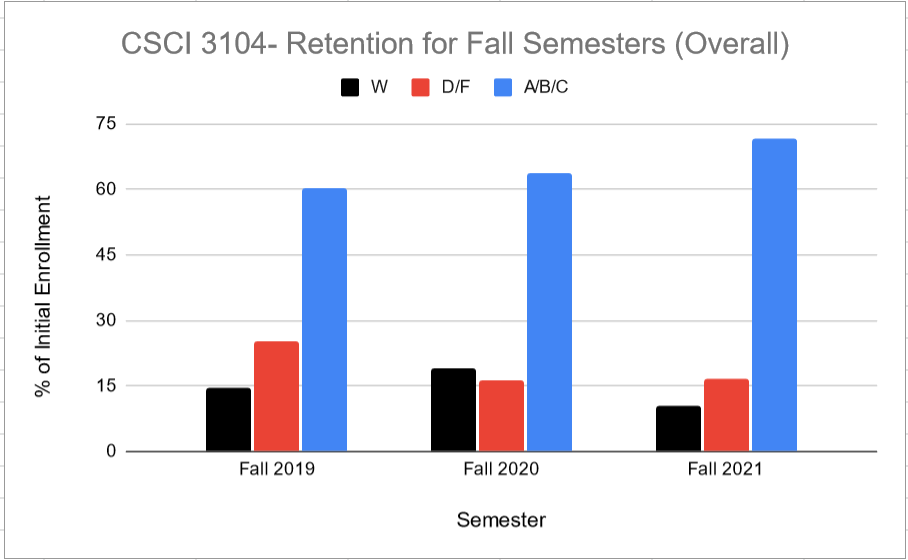} 
\caption{}
\end{subfigure}

\caption{\label{fig:retention} Retention for CSCI 3104 Fall semesters. (a) BA CS students, (b) BS CS students, (c) overall.}
\end{figure}

We highlight the impacts of our standards-based grading and curricular changes on the retention rate. See Figure~\ref{fig:retention}. This section is intended to serve as an administrative summary. For a more detailed discussion of the impacts on our changes, see Section \ref{PracticalPerformance}.

In Fall 2020, we effectively overlayed standards-based grading on top of the historical curriculum. Already, this increased the overall retention rate from 60.37\% in Fall 2019 to 64.71\%, and from 73.87\% to 83.33\% for BS CS students in the same period. While retention increased for the BA CS students, the increase was quite minimal. In Fall 2019, the retention rate for the BA CS students was 44.72\%, and this increased to 47.71\% in Fall 2020.

In Fall 2021, we refined our standards-based grading implementation, re-ordered the content, and modified how certain topics were presented (see Section \ref{sec:second}). This change was highly successful. The overall course retention rate for Fall 2021 was 71.79\%. For the BS CS students, we the retention rate was 77.17\%, which is somewhat lower than in Fall 2020, but still an improvement compared to Fall 2019. Our biggest gains were amongst the BA students, for whom the retention rate was 58.24\%. 

One concern that was raised about whether standards-based grading and our subsequent curricular changes only benefitted those who persist, and that the demands may actually cause the weaker students to drop. This concern is not entirely unfounded. While overlaying Standards-Based Grading resulted in increased pass rates amongst those that persisted to the end of the course, we saw a sharp increase in the number of BA CS students that withdrew from the course. In Fall 2019, we started the course with 270 students and ended with 231 students. In Fall 2020, we started with 272 students and ended with 220 students. So while the performance amongst the students who persisted improved in Fall 2020 compared to Fall 2019, there was an uptick in the number of students who dropped. However, with our additional curricular changes in Fall 2021, we saw a decrease in \textit{both} the D/F and withdraw rates. In particular, we began the course with 280 students in Fall 2021 and ended with 251 students. Again, we note that the retention rate in Fall 2021 was 71.79\%, which is much higher than the 64.71\% in Fall 2020 and the 60.37\% in Fall 2019.

%We include below graphical representations of the retention rates for the Fall semesters from 2019-2021, including for the class overall, the BA CS student population, and the BS CS student population.

%\newpage

\subsection{Student Performance (Semester-by-Semester)} \label{PracticalPerformance}

\begin{figure}[!htbp]
\centering

% GPA
\begin{subfigure}[b]{0.45\textwidth}
\centering
\includegraphics[scale=0.44]{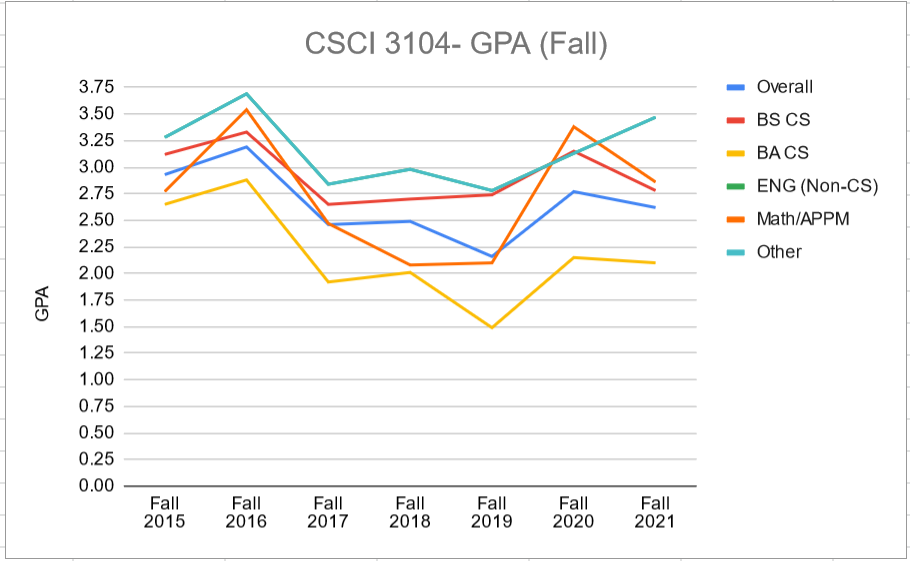}
\caption{}
\end{subfigure}
\hfill
\begin{subfigure}[b]{0.45\textwidth}
\centering
\includegraphics[scale=0.44]{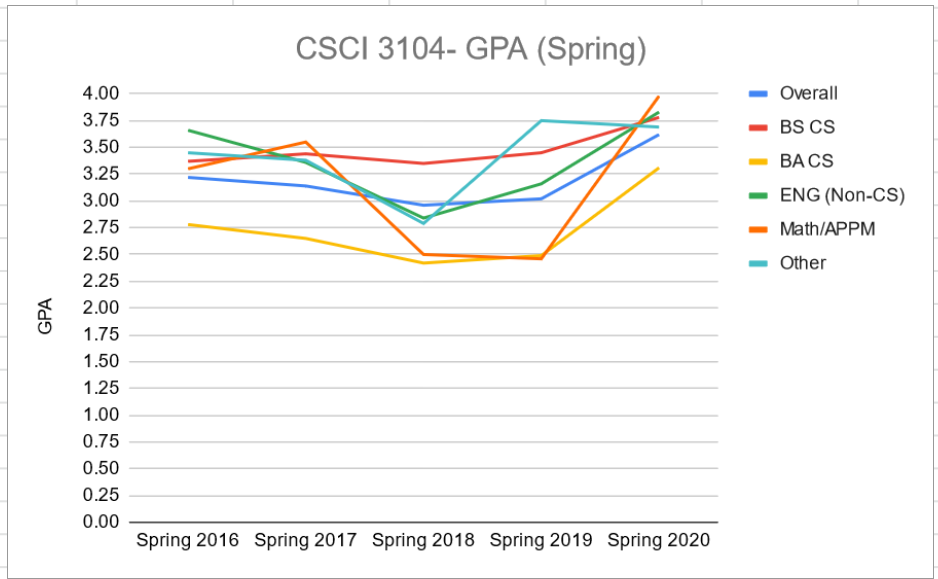}
\caption{}
\end{subfigure}

% Pass rates
\begin{subfigure}[b]{0.45\textwidth}
\centering
\includegraphics[scale=0.44]{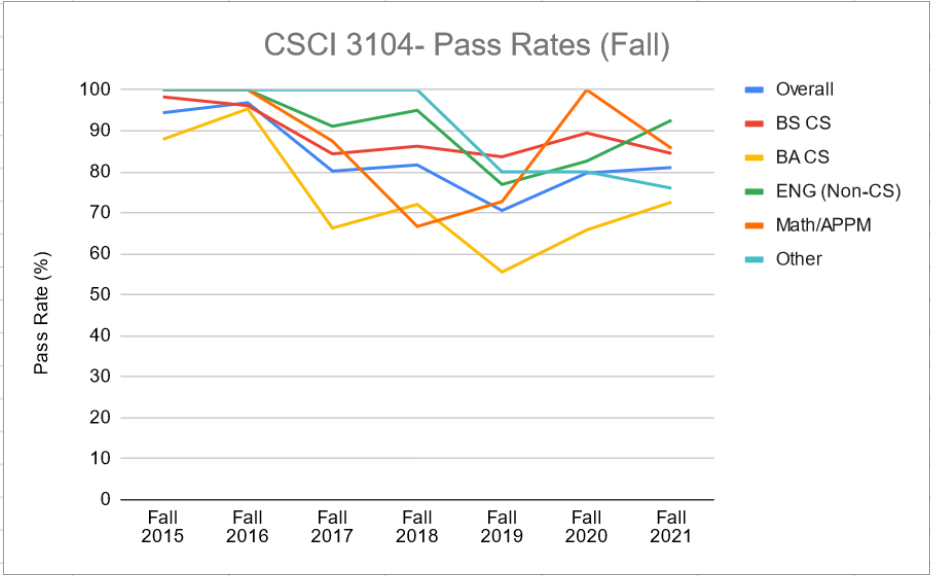} 
\caption{}
\end{subfigure}
\hfill
\begin{subfigure}[b]{0.45\textwidth}
\centering
\includegraphics[scale=0.44]{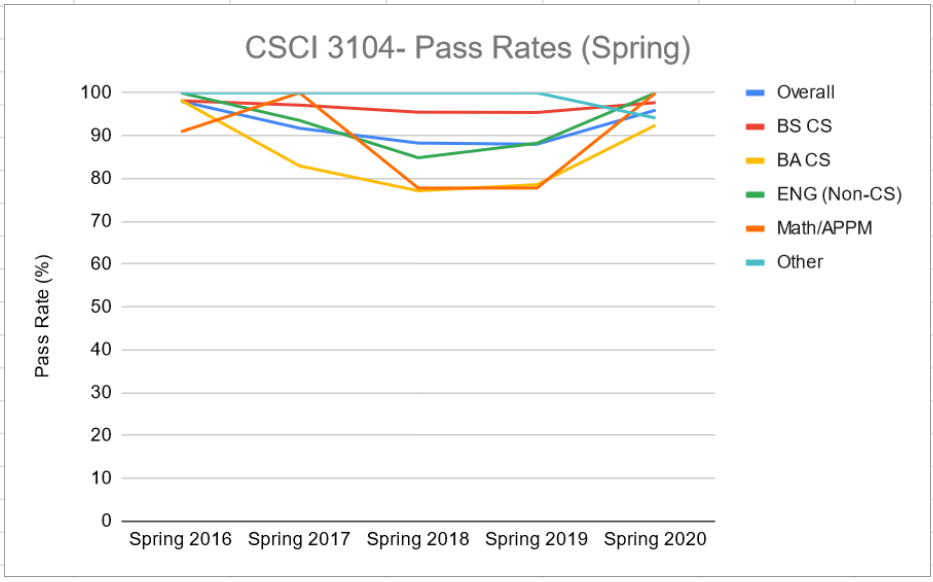} 
\caption{}
\end{subfigure}

% Retention
\begin{subfigure}[b]{0.45\textwidth}
\centering
\includegraphics[scale=0.44]{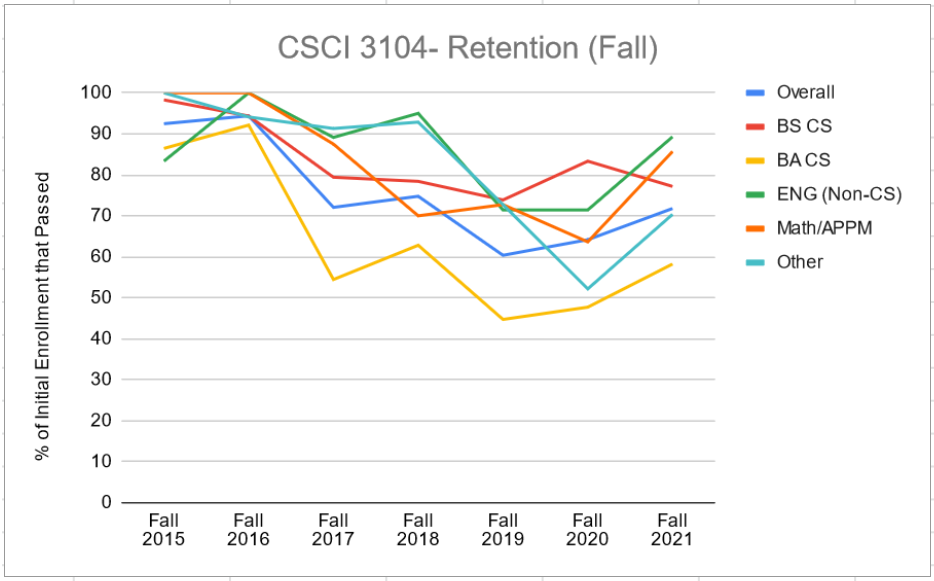} 
\caption{}
\end{subfigure}
\hfill
\begin{subfigure}[b]{0.45\textwidth}
\centering
\includegraphics[scale=0.44]{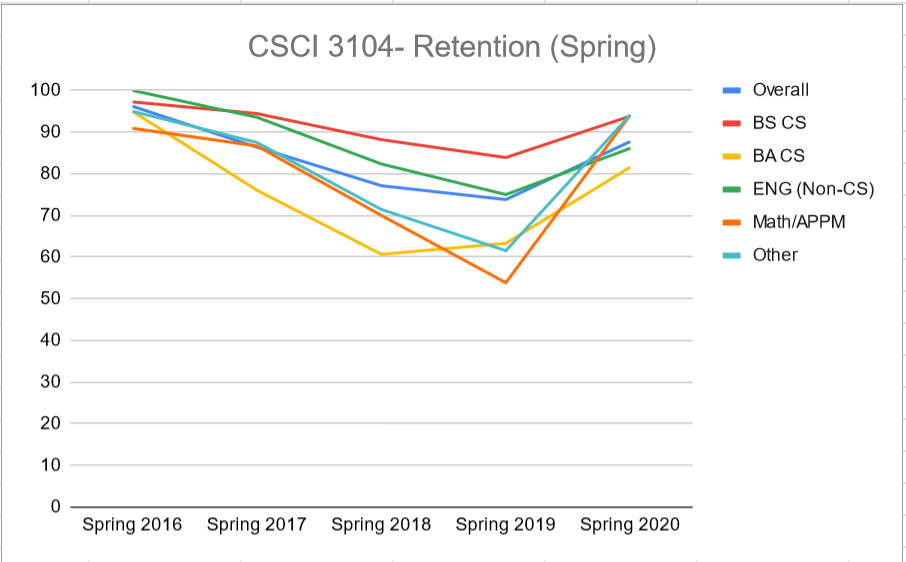} 
\caption{}
\end{subfigure}

\caption{\label{fig:semester} Semester-by-semester GPA, pass rates, and retention rates. (a) and (b): GPA for Fall and Spring semesters, respectively. (c) and (d): Pass rates for Fall and Spring semesters, resp. (e) and (f): Retention rates for Fall and Spring semesters, resp.}
\end{figure}

In this section, we discuss the impacts of standards-based grading and our curricular changes on student performance. Figure~\ref{fig:semester} summarizes GPA, pass rates, and retention rates by semester (Fall 2015 through Fall 2021) and major, divided into Fall and Spring semesters. \\

%\begin{figure}
%\begin{center}
%\includegraphics[scale=0.88]{GPA_Fall.png}
%\end{center}
%\caption{GPA for Fall Semesters}
%\end{figure}
%
%\begin{figure}
%\begin{center}
%\includegraphics[scale=0.88]{GPA_Spring.png} 
%\end{center}
%\caption{GPA for Spring Semesters}
%\end{figure}
%
%
%\begin{figure}
%\begin{center}
%\includegraphics[scale=0.88]{Pass_Rate_Fall.png} 
%\end{center}
%\caption{Pass Rates for Fall Semesters}
%\end{figure}
%
%
%\begin{figure}
%\begin{center}
%\includegraphics[scale=0.88]{Pass_Rate_Spring.png} 
%\end{center}
%\caption{Pass Rates for Spring Semesters}
%\end{figure}
%
%
%\begin{figure}
%\begin{center}
%\includegraphics[scale=0.88]{Retention_Fall_Line_Chart.png} 
%\end{center}
%\caption{Retention Rates for Fall Semesters}
%\end{figure}
%
%\begin{figure}
%\begin{center}
%\includegraphics[scale=0.88]{Retention_Spring_Line_Chart.png} 
%\end{center}
%\caption{Retention Rates for Spring Semesters}
%\end{figure}

\noindent \textbf{Fall 2020.} In the Fall 2020 semester, we implemented our standards-based grading scheme, covering the material in the traditional ordering (starting with proof by induction and asymptotics; see Section \ref{SBGAttempt1}). The students were quite vocal that they felt like they were being set up to fail, between the content ordering coupled with the demands of standards-based grading. However, there was a notable improvement from Fall 2019. For the course overall performance, the pass rate was 79.72\% (a 1.13x increase from Fall 2019), the retention rate was 64.71\% (a 1.072x increse from Fall 2019), and the GPA was 2.77 (+0.61 from Fall 2019). We also saw a significant improvement amongst the BS students, with a pass rate of 89.47\% (a 1.069x increase from Fall 2019), a retention rate of 83.33\% (a 1.128x increase from Fall 2019), and a GPA of 3.15 (+0.45 from Fall 2019). 

Amongst the BA CS population, there was a notable improvement in both the pass rate and GPA, compared to Fall 2019. The pass rate was 65.38\% (a 1.1767x increase increase from Fall 2019), and the GPA was 2.15 (+0.66 from Fall 2019). The retention rate for the BA CS students, however, was only 47.71\% (a 1.0669x increase from Fall 2019). This suggests that overlaying standards-based grading on top of the existing curriculum helped those who were either already prepared for the course or were willing to persist. However, the data suggests that standards-based grading on its own did little to address the fact that a significant number of BA CS students were still dropping the course. 

It is also worth considering whether the grade distribution for the Fall 2019 iteration was unusually low, and whether these changes would have persisted had we not implemented standards-based grading. ML worked with the course directly in Fall 2019 and Fall 2020, and believes that without standards-based grading, the pass rate in Fall 2020 would have been similar to Fall 2019.

We believe that the original ordering of the course content did little to help students, especially those with weaker backgrounds, gain traction in the course. This led us to our changes in Summer 2021 and Fall 2021 (see Section \ref{sec:second}). Under these changes, student performance continued to improve. We report grade statistics for Fall 2021. The summer sessions are not taught consistently, and so Summer 2021 does not have a base for comparison. 

\noindent \\ \textbf{Fall 2021.} For the class overall in Fall 2021, we had a pass rate of 199/253 ($78.65\%$), retention rate of 71.79\%, and a GPA of 2.62. The initial enrollments were comparable ($\sim$270-280 students) in Fall 2019, 2020, and 2021, and so the changes discussed in Section \ref{sec:second} correllated with more students passing \textit{and} fewer students dropping the course in Fall 2021. In particular, we believe that weaker students who would have dropped in previous Fall iterations persisted and passed in Fall 2021, which accounts for the decrease in GPA from Fall 2020 (though the GPA was still higher than in Fall 2019).

Somewhat surprisingly, we saw a slight drop in performance in Fall 2021 compared to Fall 2020 amongst the BS CS student population. However, the performance in Fall 2021 was still notably stronger than in Fall 2019. The pass rate was 84.48\%, the retention rate was 77.17\%, and the GPA was 2.78. While the changes we made to the curriculum, outlined in Section \ref{sec:second}, allowed us to calibrate a C- to the number of mechanical standards, we also created new standards for some of the more challenging content (e.g., designing DP algorithms, structure and consequences of P vs. NP). As a result, the expectations for grades in the B+/A-/A range were considerably higher than in previous iterations of the course. This allowed us to push our stronger students, and the BS CS student population is historically quite strong. We also note that \textit{pandemic fatigue} may have contributed to the decreased performance. Both our students, as well as students across numerous other universities, reported increased fatigue due to the COVID-19 pandemic \cite{PandemicFatigue}.

In contrast, we saw sharp improvement in performance amongst the BA student population in Fall 2021. The pass rate was 72.6\% (a 1.3067x increase from Fall 2019), the retention rate was 58.24\% (a 1.3023x increase from Fall 2019), and the GPA was 2.10 (+0.61 from Fall 2019). That is, amongst the BA students, we decreased the number of students who dropped and increased the number of students who passed. These numbers are also quite consistent with the Fall 2017 and Fall 2018 iterations of the course (see Table~\ref{table:GPA} for GPA data by semester), at which point Algorithms was not required for the BA degree. Effectively, our standards-based grading implementation and re-ordering of the curriculum counteracted negative impacts on student performance from requiring Algorithms for the BA degree.

%Below we include student performance data.

%. This contrasts with the pass rate of 163/231 (70.56\%) in Fall '19.
\begin{table}[htp!]
\begin{center}
\begin{tabular}{rccccc}
GPA & '17 & '18 & '19 & '20 & '21 \\ \hline
Fall & 2.46 & 2.49 & 2.16 & \textbf{2.77} & \textbf{2.62} \\
Spring & 3.14 & 2.96 & \textit{3.02} & \textbf{3.62} \\
Summer & & & & & \textbf{2.4}
\end{tabular}
\caption{\label{table:GPA} GPAs for the several years prior to and after implementing our changes. GPAs in \textbf{bold} follow our standards-based grading scheme. Spring '19 \textit{(italics)} followed a points-based mastery grading scheme implemented by JAG and RL that was a precursor to the changes described here. }
\end{center}
\end{table}

%The pass rates (again, we consider students who completed the course) are also of interest, with rates of 178/222 (80.18\%) in Fall '17 and 165/202 (81.68\%) in Fall '18. In Fall '19, when Algorithms became required for the BA degree, the pass rate dropped to 163/231 (70.56\%).  Under our standards-based grading scheme, we improved the pass rate to 173/222 ($77.93\%$) in Fall '20. Under the changes in Summer and Fall '21 (see Section \ref{sec:second}), the pass rates grew slightly to 51/64 ($79.68$\%) and 199/253 ($78.65\%$), respectively. The initial enrollments were comparable ($\sim$270-280 students) in Fall '19, '20, and '21, and so the changes discussed in Section \ref{sec:second} correllated with more students passing and fewer students dropping the course in Fall '21.

\section{Conclusion and Future Work}

In this paper, we introduced an application of Standards-Based Grading to our undergraduate Algorithms course. We discussed the benefits and drawbacks of our initial attempts in Spring 2020 and Fall 2020, which led to key changes in refining the list of standards, as well as the order and manner in which the content was presented. By Fall 2021, we had increased the retention and pass rates for the class overall, with stronger gains amongst the BA CS student population.

Our work has some limitations, in that some of our conclusions are drawn from anecdotal experience. For instance, our course staff have observed reduced student anxiety around grades. We have not made an attempt to rigorously ascertain the relationship between student anxiety surrounding grades and Standards-Based Grading. Additionally, our grade data and analysis do not take into account student identities. It would be of interest to analyze the relationship between Standards-Based Grading (in Algorithms) and how well students from systemically marginalized and excluded populations perform. In particular, it would be of interest to study the impacts of our Standards-Based Grading scheme on  students who often have limited time outside of class due to other obligations (e.g., work, caretaking).

It would also be of interest to study the effects of implementing Standards-Based Grading in prerequisite courses, such as our department's Discrete Math course, on student readiness for and performance in Algorithms. Furthermore, it would be interesting to analyze the impacts on retention of implementing Standards-Based Grading in both a prerequisite and the subsequent course (e.g., Discrete Math and Algorithms). A positive correlation could point the way to effectively supporting students in large classes across the undergraduate CS curriculum. As we observed in Spring 2020, implementing Standards-Based Grading at scale requires a significant investment in teaching staff (i.e., instructors, TAs, graders). We recommend considering the requisite and available teaching resources before committing to such a study.

\bibliographystyle{alphaurl}
\bibliography{references}

\end{document}